\documentclass{epl}
\def\tmre{$(TMTSF)_{2}ReO_{4}$}\,
\def\tmx{$(TMTSF)_{2}ClO_{4(1-x)}ReO_{4x}$}\,
\def\tmp{$(TMTSF)_{2}PF_{6}$}\,
\def\tmc{$(TMTSF)_{2}ClO_{4}$}\,
\,
\,
\def\tm{$(TMTSF)_{2}X$}\,
\def\R{$ReO_{4}^{-}$}  
\def\C{$ClO_{4}^{-}$}

\title{Impurity controlled Superconductivity/Spin Density Wave interplay in the
organic superconductor : $(TMTSF)_2ClO_4$}
\shorttitle{Impurity controlled Superconductivity in $(TMTSF)_2ClO_4$}
\author{N.Joo\inst{1,2} \and P.Auban-Senzier\inst{1} \and C.R.Pasquier\inst{1}\and D.J\'erome\inst{1} \and K.Bechgaard\inst{3}}
\shortauthor{Joo et-al}
\institute{
  \inst{1} Laboratoire de Physique des Solides (UMR 8502) - Universit\'e Paris-Sud, 91405,\\ Orsay, France\\
  \inst{2} Facult\'e des Sciences de Tunis, LPMC, Campus Universitaire, 1060, Tunis, Tunisie\\
	 \inst{3} H. C. $\O$rsted institute, Universitetsparken 5, DK 2100, Copenhagen, Denmark
}


\begin{document}

\maketitle

\begin{abstract}
The study of the  anion ordered \tmx \, solid solution in the limit of a low \R \,substitution level ($0\leq x\leq 17\%$) has revealed a new and interesting phase diagram. Superconductivity is drastically suppressed  as the effect of \R \, non magnetic point defects
increases following the digamma behaviour for usual  superconductors in the presence of paramagnetic impurities. Then, no long range
order can be stabilized above 0.1K in a narrow window of substitution. Finally, an insulating SDW ground state in 
\R -rich samples is rapidly stabilized  with the decrease   of the  potential strength leading to the doubling of the transverse
periodicity.  This extensive study has shown that the superconducting order parameter must change  its sign
over the  Fermi surface.

\end{abstract}

The symmetry of the superconductivity order parameter remains a
controversial question in organic superconductors where unlike the high
$T_c$ cuprates phase sensitive experiments have failed so far to provide a
clear cut answer\cite{Tsuei00}. Among the various experimental approaches used to
tackle the problem of the order parameter symmetry the sensitivity of the
superconducting state to non magnetic impurities is known to be quite
pertinent since unlike conventional \textit{s}-type superconductors,
unconventional superconductors, singlet \textit{d} \cite{Bourbonnais88} or triplet \textit{p} or \textit{f} type\cite{Abrikosov83,
Maki04, Nickel05,Kuroki05} are strongly affected by their presence.  Organic superconductors belonging
to the \tm \, family of one-dimensional conductors\cite{Jerome82}  provide a situation
which is very well adapted to the study of the role of non magnetic
impurities on the superconductivity instability. The solid solution
\tmx \, which has been the subject of an exhaustive crystallographic study\cite{Ilakovac97} represents a very good example since a
small amount of \R \, anions does not disturb the electronic structure and yet
profoundly affects the stability of the superconductivity phase of the
pure \tmc \, compound\cite{Tomic83b,Tomic86}. A preliminary study of
\tmx \, has been carried on previously\cite{Joo04} but it was restricted to
the low dilution regime namely $\leq$ 6 \%.  It was shown, plotting the value
of $T_c$ versus the residual resistivity of the normal state that the strong
suppression of $T_c$ does not support the existence in these organic
superconductors of an order parameter keeping the same sign over the
entire Fermi surface. The present work  extends the
investigation towards higher concentrations of \R anions. Furthermore,
a reinvestigation of the procedure used for the plot of $T_c$ \textit{vs} $\rho_0$ has been
performed leading to removing some flaws in the previous
interpretation\cite{Joo04}. In addition, we have studied the solid solution up to $x$=17 \%
showing that after the suppression of the SC phase there exists a
narrow window in the impurity doping where no low range order can be stabilized at low temperature. At larger
doping an insulating spin density wave (SDW) ground state  becomes stable. Sample
growing and measurements of the resistivity along the c* axis (weakest
conductivity direction) have been performed following a procedure mentioned previously\cite{Joo04}. As done
before,  the elastic electron life time of the normal
state has been characterized by the residual term derived   in a
polynomial fit of the resistivity \textit{versus} temperature in the metallic
state. We have clearly identified a temperature regime above the superconducting transition in which
precursor superconductivity is dominant over the more conventional behaviour prevailing at higher
temperatures. Consequently,\textit{at variance} with the procedure used earlier which was based on a linear
extrapolation of the temperature dependence of the resistivity below 10 K we have extracted the value of $\rho_0$
from the quadratic temperature dependence which is observed above 10 K. Hence, the previous use of a linear law to
extract the value of the residual resistivity leads to a serious underestimation of $\rho_0$. We think that
the present approach  is more appropriate to the picture
of a Fermi liquid which can be expected to hold at low temperature in these quasi
1 D conductors.  
\begin{figure}[htbp]
\centerline{\includegraphics[width=0.9\hsize]{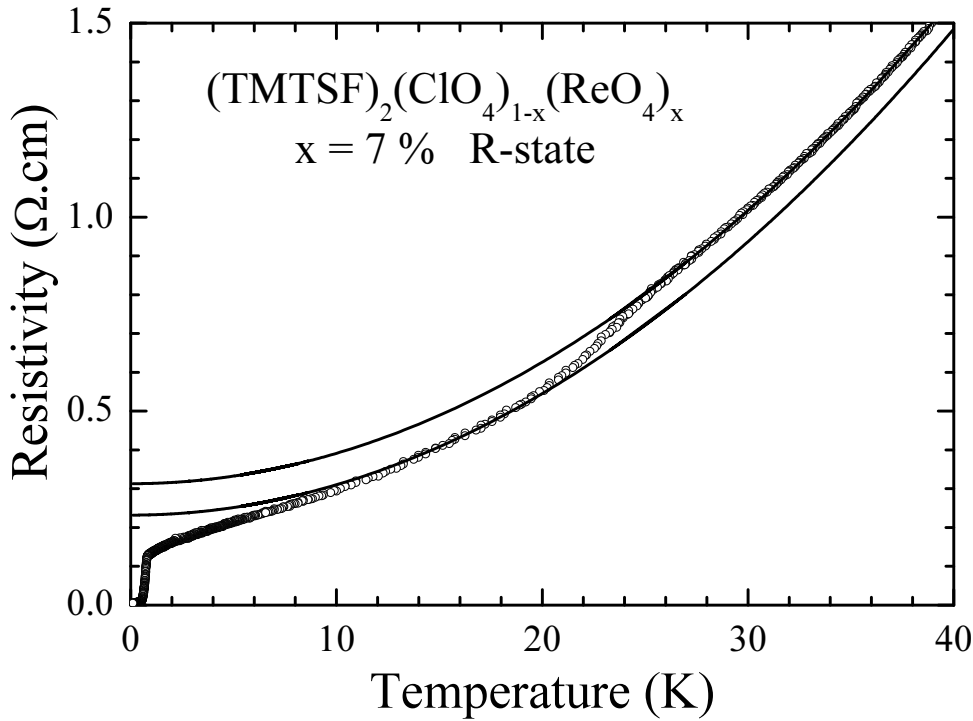}}
\caption{Temperature dependence of the resistivity in \tmx \, showing the $T^2$ law  obeyed with the same value of the prefactor A on
both sides of the anion ordering transition . The ordering amounts to a decrease of the elastic scattering. The temperature dependence
departs from the
$T^2$ behaviour in the precursor superconductivity regime below 10K or so.}
\label{f.1}
\end{figure}

Figure 1 displays the temperature dependence of the resistivity $\rho_c$
plotted against \textit{T} below 40K. The anomaly noticed at 24 K corresponds to
the well known ordering of the tetrahedral anion \C \, leading to
additional Bragg reflections\cite{Ravy86} at a wave vector (0,1/2,0) and concomitantly
to a reduction of the elastic electron scattering in the ordered
phase below $T_{AO}$. It can be noticed that a quadratic law is well obeyed
above $T_{AO}$.  This behaviour is
understandable since the \textit{T}$^2$ term is the signature of electron-electron
scattering in the Fermi liquid with the prefactor A determined by the density of states
at the Fermi level \textit{albeit} possibly enhanced by correlations. The onset of
the anion ordering amounts to a folding of the energy dispersion along
the \textit{b}-direction which should not affect N(E$\mathrm{_F}$) and hence the magnitude of
the electron-electron scattering.
Below $T_{AO}$ a quadratic law is also followed with the same
prefactor  A within the accuracy of a fit which can only be performed in a narrow temprature domain while the  constant term $\rho_0$
is diminished. In the present work it is the value of the residual resistivity $\rho_0^-$ derived from the quadratic low below $T_{AO}$
which has been used as the parameter describing the amplitude of the elastic 
scattering rate when non magnetic impurities  are added to those present initially in the pure compound.  On
figure 2 the value of $T_c$ obtained either from the onset of
superconductivity in the $\rho$(T) data or equivalently from the temperature where
H$_{c2}$ reaches a zero value has been plotted versus $\rho_0^-$ for the
 different solid solutions. For all samples, the so-called relaxed  state (R) at
low temperature  has been obtained with a slow cooling rate of -0.05K/mn
(-3K/hour) below 30 K, (in some cases  shown in the figure, faster cooling
rates have also been used).  We may notice that the amount of disorder in a sample is better characterized  by the
residual resistivity
$\rho_0^-$ than by its nominal impurity concentration given by chemistry. This can be ascribed to the impurity
distribution existing  in every batch. As displayed on fig.2, all $T_c$ data
fit with a very good accuracy the life time dependence given by the
digamma function\cite{Larkin65,Maki04},
\begin{equation}
ln(\frac{T_{c0}}{T_{c}})=\Psi(\frac{1}{2}+\frac{\alpha T_{c0}}{2\pi T_{c}})-\Psi(\frac{1}{2})
\end{equation}
with $\Psi$ being the digamma function, $\alpha = \hbar /2 \tau k_{B}T_{c0}$ the depairing parameter and $\tau$
the elastic scattering time.
ä\label{e.1}

 According to fig2 the virtual critical temperature in
the absence of any scattering amounts to 1.57 K. Pristine \tmc \, 
samples although pure enough to exhibit superconductivity between 1.20-1.30 K
are still far from  perfect purity. What is remarkable in the data
of fig.2 is the existence of a highly \R substituted sample (10\%) which
has provided different values of $T_c$ depending on the cooling rate. These data lie precisely in the regime where
the digamma function departs from the linearity \textit{versus} 1/$\tau$.
\begin{figure}[htbp]
\centerline{\includegraphics[width=0.9\hsize]{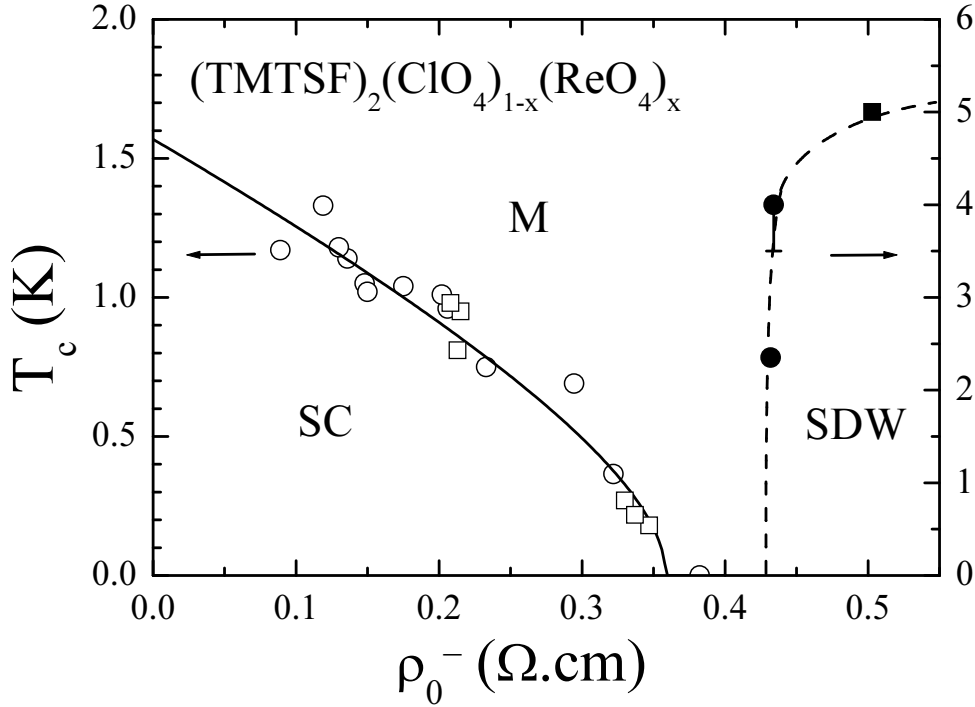}}
\caption{Phase diagram of \tmx \, governed by  non magnetic disorder. All open circles refer to the very slowly cooled samples in the
R-state with different \R contents. Open squares are  data for the same samples corresponding to slightly larger cooling rates although
keeping a metallic behaviour above $T_c$. The 10\% sample with $\rho_0^- \approx 0.32 \Omega$.cm  has provided four different $T_c$
depending on the cooling rate. One sample (8\%) did not reveal any ordering down to 0.1K. These data show that the residual
resistivity is a better characterization for the disorder than the nominal \R concentration.  Full dots (15 and 17\%) are relaxed
samples exhibiting a SDW ground state. The vertical bar is the error bar for a sample in which a maximum of the logarithmic derivative
could not be clearly identified and therefore the actual  SDW temperature should lie below 4K, the temperature of minimum
resistivity. The full square is the Q-state of a 6\% sample. The continuous line at the M-SC transition  is the  best fit of the data
 with the digamma function providing
$T_{co}$ = 1.57K. The dashed line at the M-SDW transition is a guide for the eye.}
\label{}
\end{figure}
 At higher
impurity content, ($x$=15\%) a new behaviour has been observed. The anions
 still order below $T_{AO}$ still of the order of 24K as indicated by the concomitant drop  of the residual resistivity
(although smaller than what is observed in purer samples). An upturn of
the resistivity noticed at low temperature in the R-state can be ascribed
to the onset of an insulating ground state at 2.35 K (as defined by the maximum
of the logarithmic derivative of the resistance \textit{versus} \textit{T}). 

We may relate the onset of an insulating ground
state to the occurence of SDW modulation as also checked by EPR experiments  either in  the R-state of \R - substituted
samples\cite{noteEPR} or in 
 rapidly cooled samples\cite{noteEPR, Schwenk84}.  The SDW temperature of the 15
\% sample is about half the value of $T\mathrm{_{SDW}}$ obtained  in
rapidly cooled samples (Quenched-state). The case of the 15 \% sample can be understood
in terms of the model showing how the anion ordering can lower the SDW
stability. It has been shown that the stability of a SDW ground state in the R-state of \tmc \, can be strongly
suppressed by the additional transverse potential introduced in the ordered state below $T_{AO}$\cite{Heritier84}.

What is also remarkable in the interplay between SC and SDW as
shown in fig.2 is the existence of a range of impurity concentrations illustrated by the 8\% sample in
which neither SC nor SDW phases can be stabilized down to 100 mK. Such a
behaviour is \textit{at variance} with the existence of a common border between SDW and
SC in the \textit{T-P} phase diagram of the parent superconductor \tmp.
However, as far as the latter compound is concerned impurities are not
involved in the pressure dependence of the superconducting state and the
maximum $T_c$ is located at the common border between SDW and SC. 

The good
agreement between the experimental dependence of $T_c$ \textit{versus} the elastic
scattering and the theory of the limitation of $T_c$ by non magnetic
impurities in unconventional superconductors is strongly suggestive of an
homogeneous material where the \C anions are ordered  and \R anions play only the role of
point defects. 

However, quite a different behaviour for the interplay
between SC and SDW is observed when the cooling rate is increased in a
sample of a given composition. 
\begin{figure}[htbp]
\centerline{\includegraphics[width=0.9\hsize]{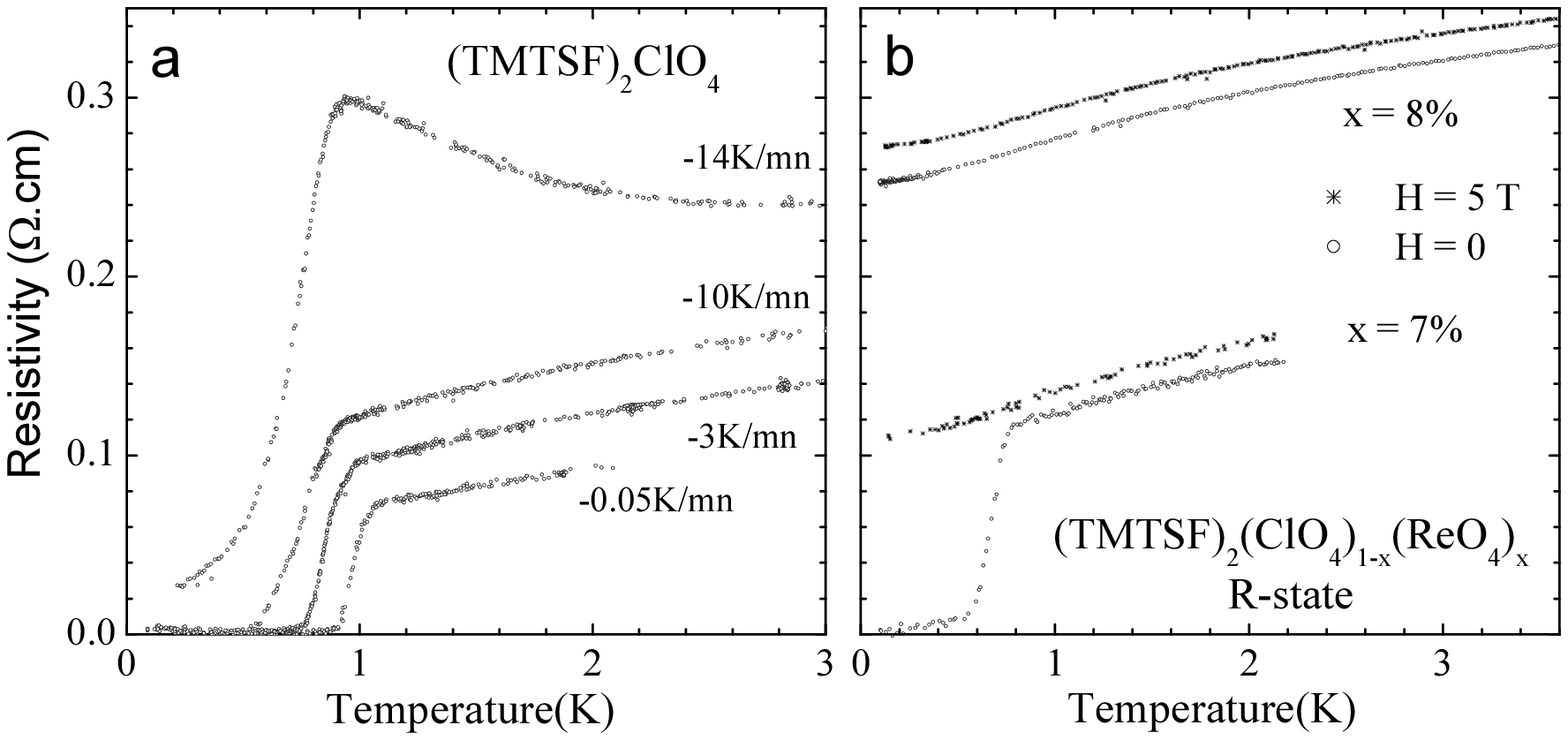}}
\caption{(a) Evolution of the resistive behaviour in pure \tmc according to the cooling rate. (b) Persistence of the presursor
regime in impure samples and under magnetic fields larger than the critical fields.}
\label{f.3}
\end{figure}
As shown in fig.3a for a pristine sample in the limit of the  low cooling rates the critical temperature
decreases gently at increasing rates with a concomitant increase of the residual resistance following   the
digamma behaviour. We can ascribe this regime to the limit where the disordered anions play the role of point
defects in an homogenous sample  and destroy the superconducting state in a way similar to the alloying effect.
However, when the cooling rate overcomes say, -3K/mn a violation of the digamma behaviour is clearly observed (not shown on fig. 2).
While $\rho_0^-$ keeps increasing with the cooling rate, $T_c$ stays constant at about 0.8K . In addition, for a rate of -14
K/mn a superconducting signature is observed at 0.8 K below a weak
upturn of the resistivity at 2.5 K.

These data can be understood
assuming the low temperature phase is becoming inhomogeneous at fast cooling rates. A
possible scenario is the occurence of macroscopic coherence  via Josephson coupling between superconducting
islands in which the anions are fairly well ordered (double sheeted Fermi surface) separated by regions in
which
 only short range order prevails in agreement with the  X-rays determination of the correlation length in quenched
samples\cite{Pouget90} . In the latter non-metallic regions
 \textit{T}$\mathrm{_{SDW}^{}}$ =~2.5 K is lower than the SDW temperature which is observed for
a fully Q-state (\textit{T}$\mathrm{_{SDW}^{Q}}$=~5 K) with a single-sheeted Fermi surface. Furthermore, on account of the loss of
the density of states at the Fermi level  below \textit{T}$\mathrm{_{SDW}^{}}$ no superconductivity can be stabilized in those 
domains which in turn remain insulating at low temperature. The evolution of the
transport properties corroborates the specific heat study which has been
performed in \tmc \,  varying the cooling rate\cite{Garoche82}. From the reduction of
the entropy involved in the SC transition it was concluded that for a
cooling rate of -10K/mn the low temperature state of the sample Ò might
contain some proportion of SDW phaseÓ. The response of the transport
properties to the cooling rate has also been observed in the \R- doped
samples but in this case the presence of point defects (\R anions)
slows down the dynamics of ordering. Consequently, in highly doped samples an inhomogeneous behaviour can more  easily be
attained even for low cooling rates .

We shall now address  the problem of the downturn of the resistivity which is clearly observed below 10K down to
the superconducting transition, (see fig.1). Such a behaviour is quite general since it had already been reported
in the discovery of superconductivity of \tmp\cite{Jerome80} and  \tmre\cite{Parkin82}  under
pressure. This downturn is robust against the presence of impurities and also against the application of a large
magnetic field, fig.3b. As far as the data of fig.3b are concerned it is important to notice that the field induced
spin density phases which should be present at low temperature in a field of 5 T //c are suppressed by impurities
although they  may possibly occur at a higher field\cite{Tsobnang94}. The theoretical basis for the downturn of the resistivity has
already been  debated in the past\cite{Schulz81} with no common consensus reached\cite{Schulz83,Kwak83} and it will be the subject of a
forthcoming publication. It was claimed that the behaviour of the resistivity could be interpreted as precursors
to the superconducting ground state with large fluctuations around a mean field transition temperature of (10-15K) within the framework
of existing band structure calculations\cite{Schulz81}. 

Without entering 
this controversy again some new experimental data of the present study should be emphasized.  Since the downturn
of the resistivity  is still observed under experimental situations where   superconductivity is suppressed  either
by  non magnetic impurities or by a magnetic field of about 8 times the critical
field along the $c$-axis (see, fig.3b) we may rule out the possibility of
paraconductive fluctuations over a temperature domain extending up to 10K
or so. Furthermore, fluctuations of a SDW contributing to a collective
mode seem unlikely because first, they would be pinned by strong
commensurability effects and second, there are no signs of magnetism in
the low temperature conducting phase of relaxed samples. We wish to
propose a possible scenario already suggested in the context of the high $T_c$
cuprate materials namely, the existence of a pseudogap with an inelastic scattering rate becoming shorter in the
pseudogap regime\cite{Timusk99}. A charge pseudogap has been identified by  tunnelling spectroscopy either in \tmc\cite{Fournel82,
Bando85} or in
 \tmp \, under pressure\cite{Fournel85} with an approximate width of $2\Delta  \approx 2-3 meV$.
Even if  the pseudogap is not due to an incoherent superconducting pair formation below a
high mean field temperature it can still be due to fluctuations of an
other type namely, antiferromagnetic fluctuations which have been
detected by NMR relaxation experiments\cite{Bourbonnais84}. Consequently, the picture of
these organic superconductors  bear some similarity with the underdoped regime of high
\textit{T}c's where both a departure of the resistivity from the high temperature
linear behaviour is observed and magnetic fluctuations have been
identified by NMR. 

The study of the solid solution \tmx \,  in the limit of a low \R \\substitution level ($x<17\%$) has revealed a new and interesting
phase diagram once the ordering of the tetrahedral anions is achieved below 24K. The insulating SDW ground state of  \R -rich samples
is quickly suppressed with an increase of the amplitude of the anion potential doubling the transverse periodicity. Then, no long range
order can be stabilized above 0.1K in a narrow window of substitution and at lower substitution level the stabiblity of a
superconducting ground state  grows as the system evolves towards pure \tmc. We are in the presence of a SDW/SC interplay of
quite a different nature than the usual interplay observed in \tmp \, under pressure. In the present case, SC and SDW are decoupled
phenomena as illustrated by the gap between these two instabilities in fig. 2 whereas the stability of superconductivity in \tmp \, 
is in close relation with  the SDW state.  The main factor governing the stability of the superconducting state in
\tmx \, and in turn its $T_c$ is the elastic electron lifetime and not the band and correlation parameters. This extensive study has
shown that the order parameter has to change  sign over the the Fermi surface.

\acknowledgments
N.~Joo acknowledges the french-tunisian cooperation CMCU (project 01/F1303) and thanks S.~Haddad, S.~Charfi-Kaddour and M.~H\'eritier for helpful
discussions and S.~Yonesawa for thermometer calibration.

\begin{thebibliography}{10}
\bibitem{noteEPR}
The SDW character of the insulating ground state stabilized in relaxed \R -
  rich samples has been established in reference\cite{Tomic83b}.
\end{thebibliography}

\end{document}